# Decoding Breast Cancer in X-ray Mammograms: A Multi-Parameter Approach Using Fractals, Multifractals, and Structural Disorder Analysis

Santanu Maity, Mousa Alrubayan, and Prabhakar Pradhan

*Department of Physics and Astronomy*
*Mississippi State University*
*Mississippi State, MS, USA, 39762*

**Abstract:** We explored the fractal and multifractal characteristics of breast mammogram micrographs to identify quantitative biomarkers associated with breast cancer progression. In addition to conventional fractal and multifractal analyses, we employed a recently developed fractal-functional distribution method, which transforms fractal measures into Gaussian distributions for more robust statistical interpretation. Given the sparsity of mammogram intensity data, we also analyzed how variations in intensity thresholds, used for binary transformations of the fractal dimension, follow unique trajectories that may serve as novel indicators of disease progression. Our findings demonstrate that fractal, multifractal, and fractal-functional parameters effectively differentiate between benign and cancerous tissue. Furthermore, the threshold-dependent behavior of intensity-based fractal measures presents distinct patterns in cancer cases. To complement these analyses, we applied the Inverse Participation Ratio (IPR) light localization technique to quantify structural disorder at the microscopic level. This multi-parametric approach, integrating spatial complexity and structural disorder metrics, offers a promising framework for enhancing the sensitivity and specificity of breast cancer detection.

## I. INTRODUCTION

Breast cancer continues to be one of the most significant causes of cancer-related mortality among women both in the United States and globally. In 2024, it is estimated that 310,720 new cases of invasive breast cancer and 42,250 related deaths were reported in women. [1–5] In addition, approximately 56,500 cases of ductal carcinoma in situ, a non-invasive precursor to breast cancer, were also documented. [6] While lung cancer remains the leading cause of cancer death in women overall, breast cancer is the primary cause of cancer death among Black and Hispanic women, highlighting the need for practical diagnostic tools across diverse populations.

X-ray mammography remains the gold standard for breast cancer screening, especially for women over 40, and is a routine part of annual health checkups. Clinicians primarily rely on visual inspection of mammograms to detect tissue abnormalities, often on a fluorescence X-ray screen, before recommending further diagnostic procedures such as biopsies. However, this method relies heavily on the subjective judgment of radiologists, making it challenging to detect early or subtle changes in tissue structure. [7–10]

The structural patterns observed in mammographic images exhibit fractal characteristics, which reflect the self-similar and irregular nature of biological tissues.[11,12] Due to sparse image intensity distribution in mammograms, these patterns display multifractal behavior. Multifractal spectrum

analysis—particularly the f(α) vs. α spectrum plot—has shown promise in quantifying malignancy-related variations in breast tissue.[13–16] Nevertheless, multifractal analysis alone has limitations: it is less sensitive to minute changes in the fractal dimension. It is challenging to translate into a fully automated numerical algorithm for diagnostic purposes, due to its non-universal functional forms.

To overcome these challenges, our study employs a comprehensive, multi-parametric approach to analyze X-ray mammograms from both benign and malignant breast tissue samples. We employ three distinct yet complementary methods: i) fractal analysis using the box-counting algorithm, ii) multifractal spectrum analysis using pixel-based f(α) vs α curves, and iii) a functional transformation of the fractal dimension that maps the distribution into Gaussian space.[11–13]. This latter approach enables more straightforward numerical comparisons through the use of the mean and standard deviation of the resulting Gaussian distribution.

In addition, we analyze tissue structural disorder using the inverse participation ratio (IPR), a light localization metric that quantifies disorder via mass density fluctuations and their spatial correlations. The structural disorder ensemble average parameter ⟨IPR⟩ is approximated as $L_{d\text{-IPR}} \sim \langle dm \rangle \cdot l_c$, where dm represents the density fluctuations and $l_c$ is their correlation length.[17–22].

Our results show consistent trends across all methods: (a) the average fractal dimension increases with disease progression, (b) the width of the multifractal spectrum broadens in malignant tissues, and (c) the Gaussian-transformed distribution shows increased mean and standard deviation (STD) values, offering better numerical resolution. Furthermore, IPR analysis confirms an increase in mass density disorder in malignant samples.

Interestingly, threshold scanning of grayscale images before binary transformation reveals a unique path in the functional fractal parameters. This evolving threshold behavior could be an additional diagnostic indicator, reflecting small-scale structural changes associated with early malignancy. Combined, these quantitative tools enhance the sensitivity and accuracy of breast cancer characterization, offering a robust, multi-perspective framework for early detection.

## II. METHODS and RESULTS

### 1. Breast X-ray mammograms

#### 1.1. Acquiring breast mammograms

Breast mammogram images were obtained from the Mendeley dataset archive, collected from the National Library of Medicine. [23] This dataset contains images of benign and malignant cancer from different parts of the breast. For our study on breast mammograms, we selected 40 mammograms from benign cases and 40 from malignant cases for further analysis.

#### 1.2. X-ray Mammogram intensity and breast mass density relation

The intensity decay in X-ray penetration is due to X-ray absorption by the soft tissues of the breast. It generally follows Beer-Lambert's law, which relates the transmission to the mass density of the X-ray radiation as it penetrates vertically (z). [24–26] In particular,

$$I(z) = I_0 \times \exp(-\mu \times z). \qquad (1)$$

I(z) is the intensity of the X-ray at the depth *z* within the tissue, $I_0$ is original intensity of the X-ray, z is the vertical penetration depth, and the constant μ is the absorption coefficient that depends on the breast tissue's intrinsic properties such as tissue absorption coefficient, mass density, fluctuations, and the overall mass density of the voxel through which the x-ray passes, in a volume of *dx×dy×z* at position (x,y). For a given intensity, the relative change of the intensity dI/I can be linked to the relative change in the mass and, in turn, the relative change in the mass density dρ/ρ:

$$dI/I \propto dm/m \propto d\rho/\rho. \qquad (2)$$

This provides the connection between mass density and X-ray intensity variations, where any intensity variations are proportional to the underlying mass density in a volume *dx×dy×z*, resulting in 2D pixels in the breast cancer mammograms.

## 2. Fractal dimension analyses of breast mammograms

### 2.1. Fractal dimension using the box-counting method

To calculate the fractal dimension of X-ray mammograms, first convert the X-ray image to binary images of 0 and 1 and then apply a box-counting algorithm. Details of the box-counting method for calculating fractal dimension are described in our previous papers and by many other authors, as referenced therein. [11,12]

We first divide the two-dimensional structures of an X-ray mammogram into many 2D boxes with different length scales for the box-counting method. Then, the average fractal dimension is calculated by fitting the slope as the length scale r is varied. The average $D_f$ value can be calculated from the slope of the regression line. $D_f = \ln(N(r))/\ln(1/r)$, which corresponds to the fractal dimension as described in [27] for details.

$$D_f = \ln(N(r))/\ln(1/r) \qquad (3)$$

Here, $D_f$ is the ensemble average fractal dimension of the samples, where N(r) represents the number of boxes needed to cover the structure. Full breast tissue is a 3D object, and the X-ray possesses very high energy and higher penetration depth. However, the final mammogram is the 2D pixelated micrographs, and the fractal dimension of the 2D mammogram will be limited to the maximum fractal dimension of $D_{f-max} = 2$.

### 2.2. Fractal dimension analyses of breast tissue and role of threshold gray scale intensity

For a standard fractal dimension calculation, the standard procedure is to choose a 50% gray-scale image as the margin for binary selection, with values of 1 (for pixels above 50%) and 0 (for pixels below 50%). [11,12,28]

X-ray mammograms are sparse and slightly biased in terms of gray-scale intensity distribution. Finding the optimal fractal dimension using a grayscale threshold value would be an interesting approach for calculating the fractal dimension. The varying threshold value indicates that the ensemble-averaged optimal fractal dimension appears at different threshold values, and there is a notable difference between benign and malignant cases.

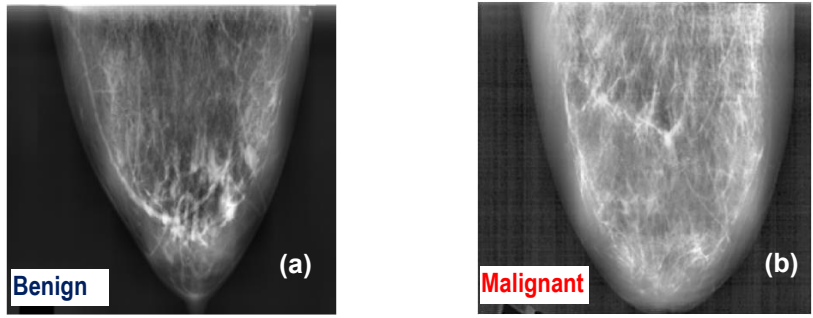

**Figure 1.** Representative breast X-ray mammograms for: **(a)** Benign and **(b)** Malignant breast mammograms.

## 2.3. Variation of Mean($D_f$) and STD($D_f$) with the gray scale threshold intensity percentage

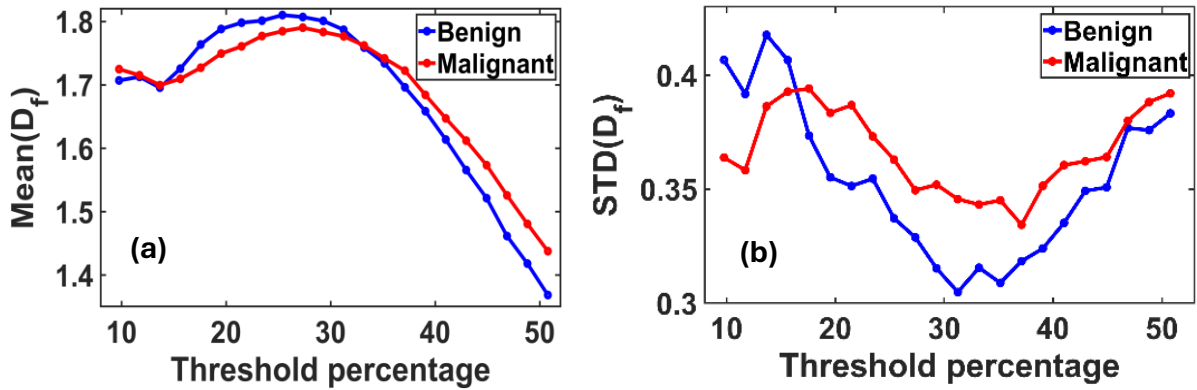

**Figure 2. (a)** Variation of the mean fractal dimension ($D_f$) and **(b)** standard deviation STD ($D_f$) as functions of threshold pixel percentage, ranging from 10% to 60% of the grayscale (default threshold is at 50%). Both plots reveal distinct peak patterns that differ between benign and malignant breast tissues. These differences indicate an optimal threshold value at which the separation in $D_f$ characteristics between the two classes is most pronounced.

The fractal dimension, $D_f$, was calculated using the box-counting method, as described in Eq. (1).

In Figure 2, we present the variation in the ensemble-averaged mean fractal dimension, Mean($D_f$), and its standard deviation, STD($D_f$), as functions of the threshold pixel values. A noticeable distinction between benign and malignant cases emerges when threshold values exceed 33% (corresponding to a grayscale value of 85 on a 0–255 scale). For the ensemble analysis, we evaluated 40 mammograms each from benign (control) and malignant cases.

Figure 2(a) illustrates the maximum difference in the mean fractal dimension between benign and malignant tissues, which occurs at an optimal threshold of 50.78%. At this threshold, the Mean($D_f$) values are 1.3687 for benign and 1.4379 for malignant tissues.

Figure 2(b) shows the corresponding variation in STD($D_f$), with the most significant difference observed at a threshold of approximately 31%. At this point, the standard deviation of $D_f$ is 0.3049 for benign and 0.3457 for malignant tissues.

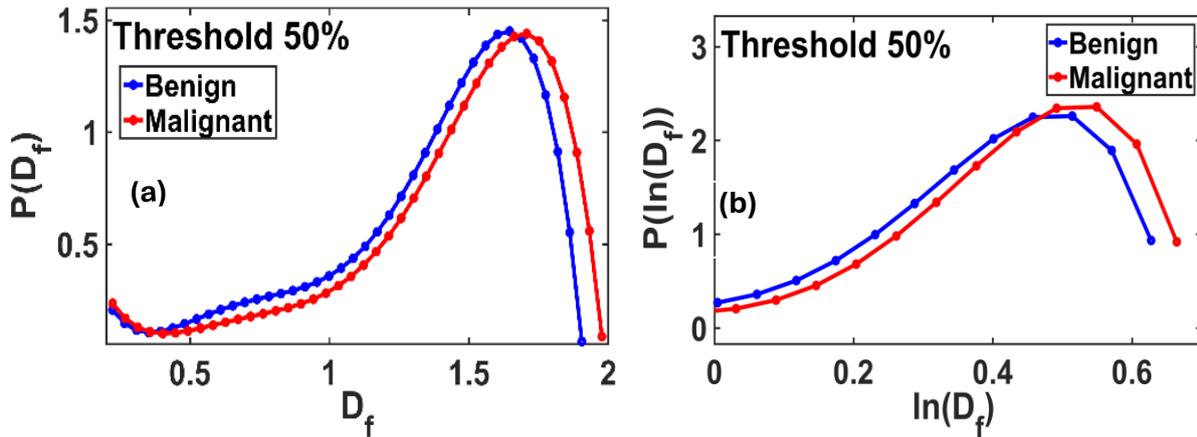

**Figure 3**. $D_f$ and $\ln(D_f)$ distributions at a standard threshold 50% of the gray scale: **(a)** The fractal $P(D_f)$ vs. $D_f$ plot shows a non-Gaussian distribution, with a slightly extended-tailed distribution for benign and malignant. **(b)** The fractal $P(\ln(D_f))$ vs $D_f$ also shows a long-tailed non-Gaussian distribution for benign and malignant.

Due to the highly fluctuating nature of the density, general and lognormal distributions of $D_f$ were checked.

These findings suggest that the mean and variability of fractal dimensions serve as distinguishing metrics for identifying malignancy in breast tissue, particularly within specific threshold ranges.

Figure 3. (a) demonstrates that the $P(D_f)$ vs $D_f$ plot exhibits a long-tailed, non-Gaussian distribution. Figure 3. (b) demonstrates that the $P(\ln(D_f))$ vs $\ln(D_f)$ also has a non-Gaussian, tailed distribution. This implies that the $D_f$ function does not follow a normal/Gaussian or lognormal distribution.

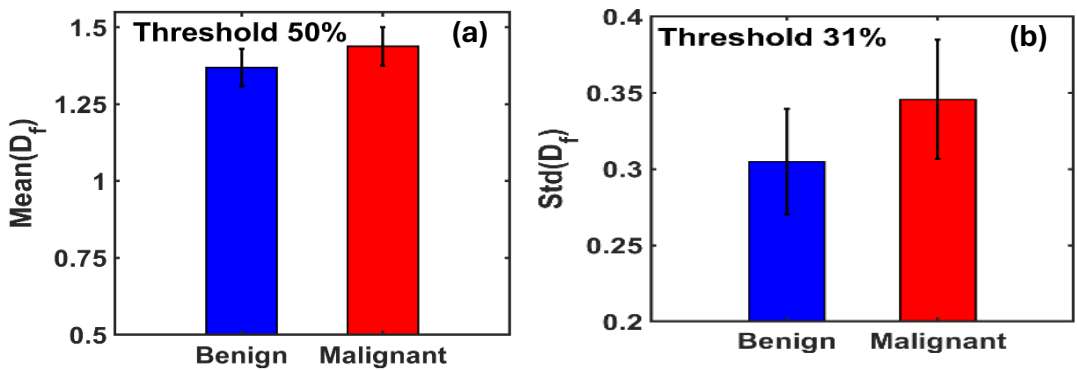

**Figure 4. (a)** Bar graphs of the mean fractal dimension (Mean($D_f$)) for benign and malignant breast tissues indicate that malignant tissues exhibit a higher mean $D_f$ value, with a percentage increase of 5.06% compared to benign tissues, at ~50% threshold.
**(b)** Bar graphs of the standard deviation of fractal dimension (STD($D_f$)) show that malignant tissues also have a higher STD($D_f$), with a 13.38% increase relative to benign tissues at 31% threshold.

## 3.2 Multifractal Analysis of Breast Cancer Mammograms

### 3.2 Multifractal Formalism of Breast Cancer Mammograms

Structural analysis of breast tissues can be performed using conventional box-counting fractal methods, which quantify the complexity of self-similar patterns. However, due to the inherent sparsity and heterogeneity of breast tissue, especially in cases of malignancy, these structures often exhibit multifractal behavior, necessitating more advanced analytical techniques.

Multifractal analysis is performed using the probabilistic box-counting method, where mammogram micrographs of size L × L are subdivided into smaller boxes of scale ε × ε. For each box, the pixel-based mass probability is calculated as [13,14,29]

$P_{\varepsilon,i} = N_\varepsilon(i)/N_{total}$,

$N_{\varepsilon(i)}$ is the number of pixels in the *i*th box, and $N_{total}$ is the total number of pixels. This distribution follows a power law, $P_{\varepsilon,i} \sim \varepsilon^{\alpha_i}$, and allows computation of the multifractal spectrum via:

$\mu_{i(Q,\varepsilon)} = P^Q_{i(Q,\varepsilon)} / \sum_{i=1:n\varepsilon}(P^Q_{i(Q,\varepsilon)})$,

And one gets the spectral relation:

$$f(\alpha_Q) = Q \times \alpha_Q - \tau_Q = \sum_{i=1}^{n\varepsilon}(\mu_{i(Q\varepsilon)} \times \ln(\mu_{i(Q\varepsilon)})) / \ln \varepsilon. \qquad (4)$$

This framework captures how the structure changes with varying *Q* (typically scanned from -10 to +10), enhancing sensitivity to sparsity and heterogeneity across different scales.
We applied this multifractal analysis to mammographic micrographs from benign and malignant breast tissues.

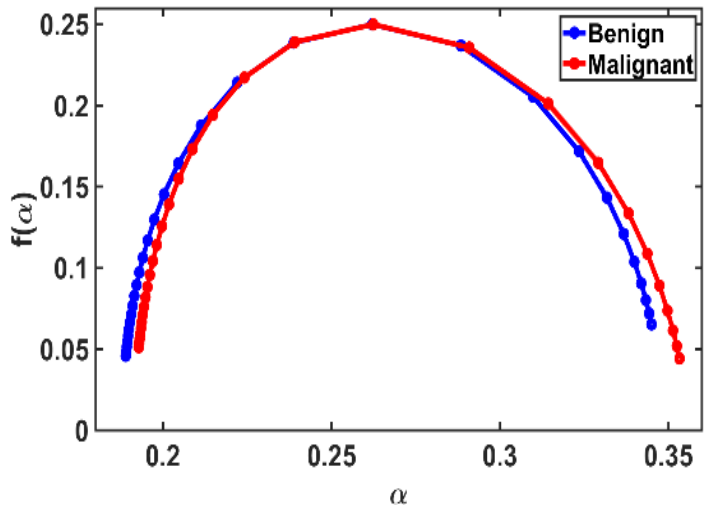

**Figure 5:** Multifractal spectrum f(α) vs. α plot for breast mammogram benign and malignant samples for each (N=40). This spectrum distribution reveals minor differences between malignant and benign tumors.

Figure 5 shows the average multifractal spectrum f(α) vs. α plot, which exhibits a slightly broader spread in malignant samples, indicating greater heterogeneity and structural irregularity. In contrast, benign tissues present a narrower spectrum, centered around fewer singularities, suggesting more uniform structures.

This observation aligns with multifractal theory, which predicts that tissues with localized mass concentrations—standard in malignancies—yield broader spectra. Conversely, more homogeneous or healthy tissues result in flatter, narrower spectral profiles. [16,30]

We extracted key multifractal spectrum parameters, including the spectrum bandwidth ($\Delta\alpha$), spectrum height ($\Delta f$), maximum and minimum singularity exponents ($\alpha_{max}$ and $\alpha_{min}$), and peak and minimum singularity strengths ($f_{max}$ and $f_{min}$), which collectively define the multifractality strength. To assess the statistical significance of these parameters in distinguishing between benign and malignant tissues, we performed a Student's *t*-test.

To quantify these differences, we extracted key multifractal parameters in the following table:

| Parameter | Benign | Malignant |
|---|---|---|
| $\alpha_{min}$ | 0.1889 | 0.1928 |
| $\alpha_{max}$ | 0.3450 | 0.3533 |
| $\Delta\alpha$ | 0.1560 | 0.1605 |
| $f_{min}$ | 0.024014 | 0.016556 |
| $f_{max}$ | 0.250000 | 0.250000 |
| $\Delta f$ | 0.2260 | 0.2334 |

**Table 1**: Comparison of different parameters, multifractal spectrum parameters of malignant breast tissue relative to benign. It can be observed that the parameter values do not change significantly.

These results support previous findings, indicating that malignant tissues exhibit larger multifractal spectral bandwidth and height, reflecting increased disorder and complexity. Although some differences are subtle and statistical significance (as determined by Student's t-test) may not be consistently strong, the observed trends provide valuable diagnostic insights.

## 4. Functional transform approach of fractal dimension $D_f$ to a new distribution

### 4.1. Functional transformation of the fractal dimension and Gaussian distribution of the transformed fractal dimension

To further enhance diagnostic resolution, we introduced a functional transformation of the fractal dimension in this context, mapping the fractal measure into a Gaussian space. This transformation facilitates statistical interpretation by focusing on changes in the mean and standard deviation of the transformed distributions.

In parallel, we also employed Inverse Participation Ratio (IPR) analysis, quantifying structural disorder through light localization metrics. Combined, these methods yield quantitative biomarkers that distinguish between benign and malignant states with improved sensitivity.

This integrated approach, combining fractal, multifractal, functional transformation, and IPR analyses, presents a robust framework for identifying subtle early-stage structural changes in breast cancer, with strong potential for clinical application.

It can be emphasized that the multifractality test indicates the samples' multifractal nature; however, quantifying the number from the *f(α)* vs α spectrum is usually challenging due to its non-universal nature.

From multifractal analysis, we see that the spectrum spreads more from control to disease.. We can use it as a diagnostic tool for cancer detection. Still, quantifying the parameters for the multifractal spectrum is challenging because the breast tissue is slightly multifractal. Using the multifractal spectrum, we can distinguish cancer stages by visual inspection and non-universal parametrization or challenging parameter quantification. [16,30]

Here, we introduced a new fractal dimension variable for quantification by applying a pointwise functional transformation to the fractal dimension $D_f$. We recently introduced and analyzed a new distribution function associated with direct fractal-functional transformations to make them a normal/Gaussian distribution, thereby facilitating the better handling of relative biomarker parameters.

To examine this aspect, we first calculate the sample's fractal dimension distribution; most of the time, it exhibits a broad tail distribution. We then check the lognormal distribution of the fractal distribution. It will be demonstrated that these do not follow a Gaussian distribution. As we perform calculations in the 2D lattice pixels, this translates to a closed fractal dimension value for $D_f \geq 2$. Therefore, obtaining Gaussian is challenging. We apply a functional transformation to each fractal point to broaden/unfold the distribution. For this, we further expand the tail using an unfolding transformation. The functional transformation is the following [13]:

$$D_{tf} = \frac{D_f}{D_{f-max} - D_f}. \qquad (5)$$

Where $D_{f-max}=2$ is the maximum fractal dimension of the micrograph in two dimensions. This new function provides a distribution with a long tail and a broader range. It has been demonstrated that the new function's log-normal distribution has a Gaussian form, or the $P(\ln(D_{tf}))$ distribution form is a Gaussian, whose mean and standard deviation follow a better pattern to handle an increasing pattern or parameter value.

## 4.2. Gray-scale threshold variation for the binary fractal dimension calculations and its effects on the different parameters

In standard fractal dimension analysis, grayscale mammogram images are typically converted to binary form using a fixed threshold, commonly assigning a value of 0 to pixels with a grayscale intensity below 50% and 1 to those above [27]. Our study expanded this approach by systematically varying the grayscale threshold to explore its potential as a novel biomarker for detecting structural changes associated with breast cancer progression.

We found that the trajectory of changes in fractal parameters as a function of grayscale threshold reveals important diagnostic information. Specifically, the threshold value at which maximum or optimal tissue parameters occur differs across tissue types: cancerous tissues reach this optimum earlier than benign cases. In contrast, benign tissues consistently show lower response levels. This trend reflects the progressive nature of carcinogenesis and provides a surprisingly effective method for tracking disease development.

This introduces the grayscale threshold variation as a quantitative diagnostic parameter, offering a new dimension for monitoring breast cancer progression. Our results section identified a specific threshold range during binary conversion that consistently produced statistically significant distinctions between benign and malignant tissues.

Each threshold adjustment modifies the binary image's mass density distribution, which affects the fractal dimension. By analyzing these changes, we establish a clear correlation between mass density variation, fractal geometry, and the malignancy stage, thereby further supporting the diagnostic utility of this method.

### 4.3. Change of parametric distribution with the variation of the gray-scaled threshold for control and malignant

We analyzed how the Mean ($ln(D_{tf})$) and standard deviation ($STD(ln(D_{tf}))$) vary with changes in the grayscale threshold used for binary image conversion. As shown in the figure, a clear distinction between benign and malignant tissues emerges at specific threshold levels: the Mean($ln(D_{tf})$) shows the most significant separation at thresholds above 50% of the maximum grayscale value, while the $STD(ln(D_{tf}))$ demonstrates the most significant differentiation around 40% of the grayscale range.

### 4.4. Lognormal distribution of $D_{ft}$ or $P(ln(D_{ft}))$ as Gaussian

Figures 6(a) and 6(b) illustrate the variation in the mean ($ln(D_{tf})$) and standard deviation ($ln(D_{tf})$), respectively, showing a consistent increasing trend from benign to malignant breast tissues. These transformed fractal parameters follow a Gaussian-like distribution, enabling straightforward quantification using the mean and standard deviation.

The analysis accounts for variations in grayscale thresholding during binary image conversion, a crucial step in calculating fractal dimensions using the box-counting method. While the default threshold is typically set at 50% of the grayscale range, our findings reveal that maximum changes in fractal dimension occur at thresholds different from this default, due to the inherent grayscale bias toward lower intensities in thick or whole-breast mammograms.

This deviation leads to identifying an optimal threshold at which the fractal dimension reflects maximum fractality—a concept we define as the "virtual fractal." The progressive increase in Mean($ln(D_{tf})$) and STD($ln(D_{tf})$) with disease severity suggests that these parameters serve as robust quantitative biomarkers for disease severity.

Importantly, this functional transformation approach offers a novel, cost-effective, and scalable early breast cancer detection, potentially improving diagnostic sensitivity and enabling broader clinical applications.

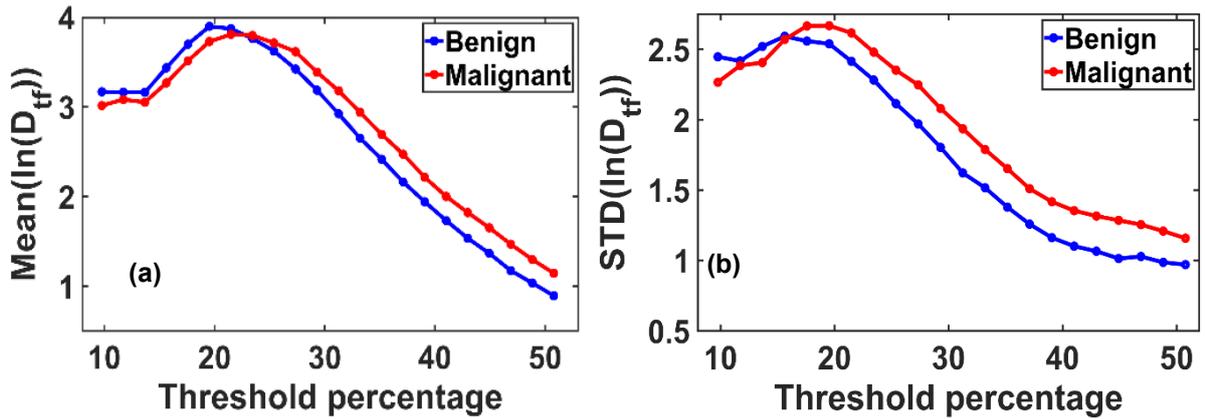

**Figure 6. (a)** Variation in the mean of the logarithmically transformed functional fractal dimension, Mean $(\ln(D_{tf}))$, and **(b)** its standard deviation, STD$(\ln(D_{tf}))$, as a function of pixel intensity threshold levels. Thresholds range from 10% to 60% of the gray scale value. The plots reveal distinct peak behaviors and threshold-dependent trends that differ between benign and malignant breast tissues. These results indicate the existence of an optimal threshold value that maximizes the contrast in fractal dimension characteristics, offering a potential criterion for diagnostic differentiation.

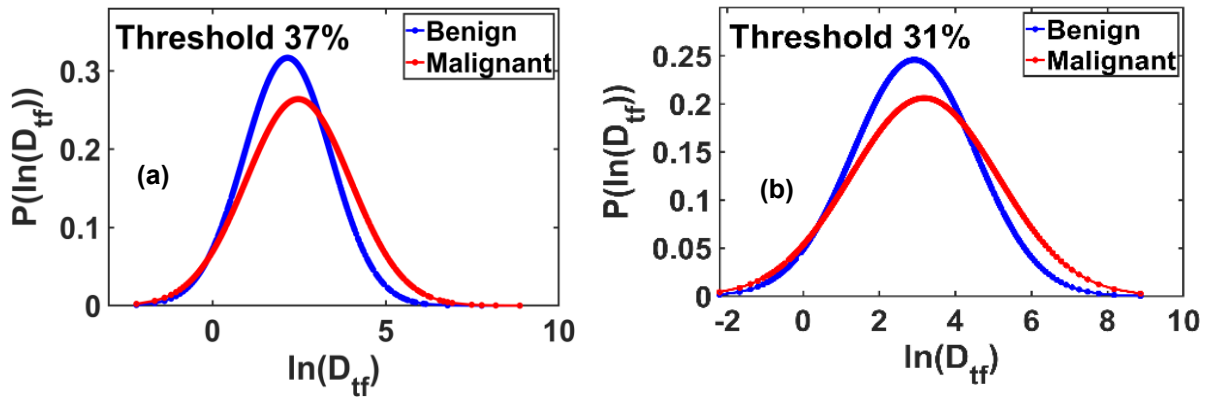

**Figure 7. (a)** Probability distribution plots of $\ln(D_{tf})$ demonstrate Gaussian behavior for benign and malignant breast tissues, with Gaussian fits yielding chi-square goodness-of-fit values exceeding 90%. The most pronounced separation in the mean $\ln(Dtf)$ values between benign and malignant tissues is observed at a threshold intensity of 37% of the maximum grayscale value.
**(b)** Similar Gaussian distributions are observed in the maximum standard deviation of $\ln(D_{tf})$, with chi-square fitting scores again exceeding 90%. The maximum differentiation in standard deviation between benign and malignant groups occurs at a threshold value of 31% of the pixel maximum. *These findings suggest threshold-dependent sensitivity of fractal dimension distributions as potential diagnostic biomarkers.*

Figure 7(a) presents the probability distribution $P(\ln(D_{tf}))$ versus $\ln(D_{tf})$, revealing a well-defined Gaussian profile with chi-square goodness-of-fit scores exceeding 90%. The most pronounced separation between benign and malignant samples in the mean values of the $\ln(Dtf)$ distribution was observed at a threshold of 37% of the grayscale range.

Similarly, Figure 7(b) shows the $P(\ln(D_{tf}))$ versus $\ln(D_{tf})$ plot corresponding to the maximum standard deviation values. This distribution also fits a Gaussian model with chi-square scores above 90%. The clearest distinction between benign and malignant tissues in terms of the standard deviation of ln(Dtf) occurred at a threshold of 31%.

These results highlight the diagnostic potential of threshold-dependent functional fractal parameters, particularly when mapped into Gaussian space.

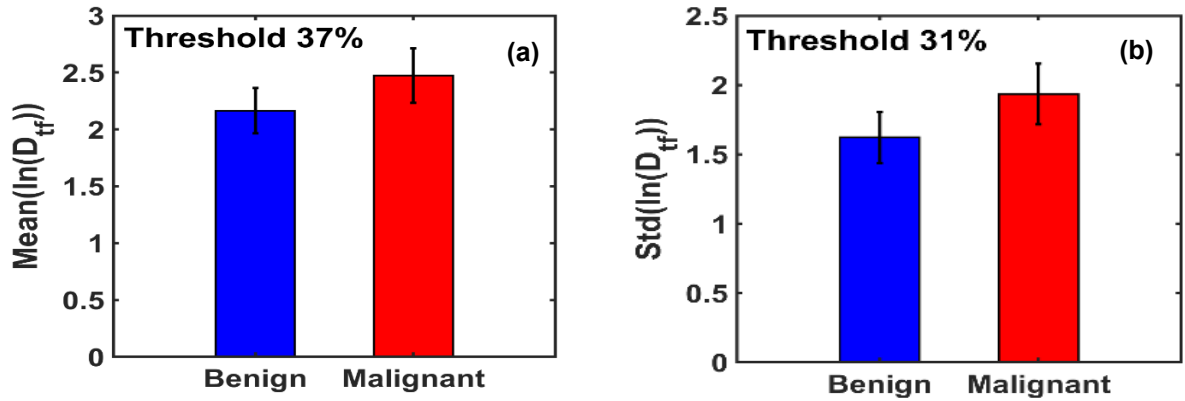

**Figure 8. (a)** Bar graph of the mean ln(Dtf) values shows that malignant breast tissues exhibit a 14.33% higher mean than benign tissues. **(b)** Bar graph of the standard deviation of ln(Dtf) reveals a 19.32% increase in malignant tissues relative to benign, highlighting greater structural variability.

Figure 8(a) illustrates that at the optimal threshold of 37%, the mean ln($D_{tf}$) values for benign and malignant breast tissues are 2.1635 and 2.4735, respectively, indicating a clear distinction between the two groups.

Figure 8(b) displays the maximum difference in the standard deviation of ln($D_{tf}$) between benign and malignant tissues, observed at the optimal threshold of 31.25%. At this threshold, the standard deviation values for benign and malignant tissues are 1.6222 and 1.9357, respectively, further supporting the sensitivity of this metric in differentiating tissue states.

## 5. Spatial structural disorder of X-ray mammogram

### 5.1 Light Localization Analysis of Breast Tissue Using the Inverse Participation Ratio Technique (IPR)

Recently adapted for biomedical imaging, the Inverse Participation Ratio (IPR) technique provides a quantitative measure of virtual light localization in breast tissue based on 2D mammographic micrographs. This technique provides a single-parameter characterization of structural disorder, and its utility is well established in mesoscopic condensed matter physics. A low mean IPR (⟨IPR⟩) value in biological tissues corresponds to reduced mass density fluctuations, suggesting relative structural uniformity. Conversely, a high ⟨IPR⟩ value indicates pronounced heterogeneity, reflecting a non-uniform spatial distribution of mass or pixel intensities and increased local disorder. [17–22]

A direct correlation exists between variations in tissue mass density and the intensity of X-ray radiation transmission through the sample. Breast tissue, being inherently heterogeneous and structurally disordered, exhibits variations in X-ray transmission intensity due to its internal composition. These variations arise from the spatial accumulation of biomolecular constituents, such as DNA, RNA, lipids, and heterochromatin, which contribute to localized changes in mass density and refractive index. The mass density ρ(x, y) within a given voxel is directly related to the refractive index n(x, y), which, in turn, governs the transmitted X-ray intensity I(x, y). This relationship is given by: [17–22]

$$I(x, y) \propto n(x, y) \propto \rho(x, y). \qquad (6)$$

Correspondingly, the optical potential $\varepsilon_i(x, y)$, which arises from refractive index perturbations, can be expressed as:

$$\varepsilon_i(x, y) = \frac{dn(x,y)}{n_0} \propto \frac{dI(x,y)}{I_0}. \qquad (7)$$

In Equations (6) and (7), $I(x,y)=I_0(x,y)+dI(x,y)$ denotes the X-ray intensity at voxel position (x, y), ρ(x,y) is the local mass density, and $n(x,y)=n_0+dn(x,y)$ represents the effective refractive index derived from density fluctuations.

Given the spatial refractive index map, the light localization properties of the tissue can be modeled using Anderson's tight-binding model (TBM), a pseudo-parametric formalism developed initially in the context of disordered electronic transport systems [31,32]. The TBM Hamiltonian is defined as:

$$H = \sum \varepsilon_i |i><i| + t\sum \langle ij \rangle (|i><j| + |i><j|) \qquad (8)$$

Here, $\varepsilon_i$ denotes the optical potential energy at site *i*, |i⟩ and |j⟩ are the optical wavefunctions at sites *i* and *j*, respectively, and ⟨i,j⟩ indicates nearest-neighbor coupling with a hopping energy parameter *t*.

The mean IPR, which quantifies the degree of light localization, is computed as: [33,34]

$$\langle IPR \rangle_{L \times L} = \frac{1}{N} \sum_{i=1}^{N} \int_0^L \int_0^L E_i^4(x, y) dx dy \qquad (9)$$

In this expression, $E_i(x,y)$ is the eigenfunction of the Hamiltonian at the *i*th lattice site, and $N=(L/dx)^2$ represents the total number of lattice points in a region of size L×L.

Our previous studies have demonstrated that both the mean IPR and its standard deviation STD(IPR) are proportional to the degree of structural disorder, characterized by the disorder strength parameter $L_d$, defined as:

$$L_d = \langle dn \rangle \cdot l_c \qquad (10)$$

Consequently, the following proportionalities can be established:

$$\langle IPR(L) \rangle_{L \times L} \propto L_{d-IPR} \sim <dn> \times l_c \qquad (11)$$

$$STD(IPR) \propto L_{d-IPR} \sim <dn> \times l_c \qquad (12)$$

These relationships establish IPR-based analysis as a sensitive and robust method for quantifying microscale structural disorder in breast tissue, offering potential utility in diagnostic imaging and tissue characterization.

## 5.2. IPR Analysis of Benign and Malignant Breast Tissue Mammogram

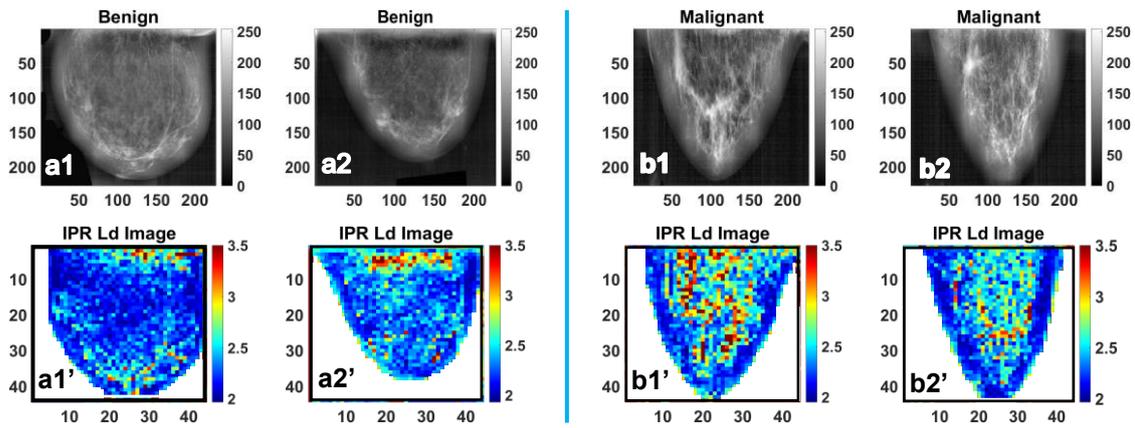

**Figure 9. (a1)** and **(a2)** are the representative gray colormap images of benign breast tissue, whereas **(a1')** and **(a2')** are their respective <IPR> ~ $L_{d\text{-}IPR}$ color maps. Also, (b1) and (b2) are the representative colormap images of the Malignant Breast tissue, whereas (b1') and (b2') are their respective <IPR> ~ $L_{d\text{-}IPR}$ color map.

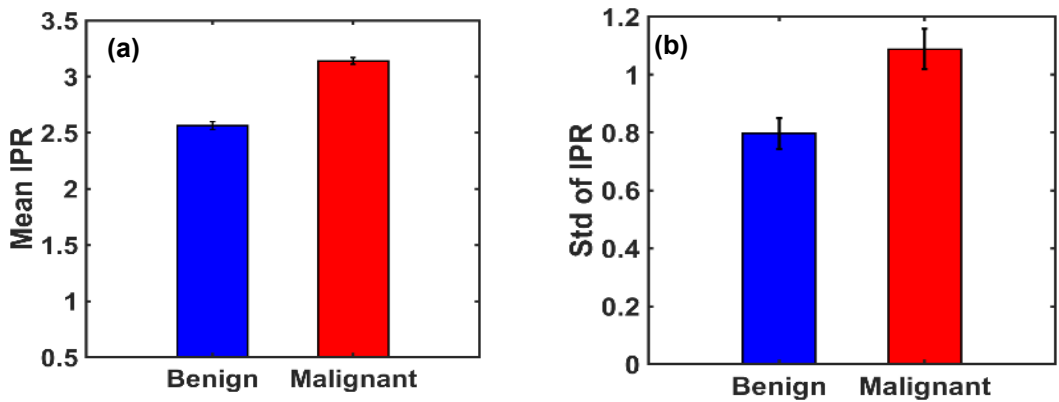

**Figure 10. (a)** and **(b)** are the bar graphs of IPR mean and IPR standard deviation of benign and malignant breast tissues. The mean IPR value is higher in malignant breast tissues than in benign breast tissues by a percentage change of *22.4634%*. STD of IPR value is higher in malignant breast tissues than benign by a percentage change of *36.619%*.

For the IPR analysis, breast tissue samples were categorized into two groups: benign and malignant. A total of 80 high-quality mammographic images were selected, comprising 40 images from each of the two groups. Representative X-ray mammogram images and their corresponding IPR maps for benign breast tissue are shown in Fig. 9.(a1)–(a2), while their corresponding IPR images are shown in Fig. 9(a1')-(a2'). Similarly, representative images of malignant tissue are presented in Fig. 9(b1)–(b2), and corresponding IPR images are shown in Fig. 9(b1')- (b2').

To facilitate quantitative comparison, statistical analysis was performed across the ensembles. The Mean IPR values (⟨IPR⟩) and their standard deviations (STD(IPR)) for each group are depicted in Fig. 10(a) and Fig. 10(b), respectively. Before analysis, all images were preprocessed by cropping to isolate the breast region, thereby minimizing background artifacts.

The analysis revealed a substantial increase in both the mean and variability of IPR in malignant tissues compared to benign. Specifically, the percentage change in ⟨IPR⟩ was approximately 22.46%, while the change in standard deviation reached 36.62%, indicating significantly higher structural disorder in malignant breast tissues.

## III. DISCUSSION AND CONCLUSIONS

In this study, we presented a comprehensive, multi-parameter analytical framework for quantitatively assessing breast cancer using X-ray mammographic micrographs sourced from publicly available radiological archives. By leveraging Beer's law, we established a direct link between X-ray transmission intensity and local mass density variations within breast tissues—variations intrinsically related to the refractive index and, consequently, the structural organization of tissue components.

Our approach integrates fractal and multifractal analyses to quantify spatial heterogeneity in mass density. The functional transformation of fractal dimension distributions followed a log-normal behavior, with the logarithmic transformation yielding Gaussian-distributed values. Notably, the mean and standard deviation of the logarithmically transformed fractal dimension ($\ln(D_{tf})$) demonstrated a consistent increase with cancer progression, providing robust and quantifiable biomarkers for malignancy.

Additionally, using the Inverse Participation Ratio (IPR) technique, adapted from mesoscopic physics, enabled sensitive detection of structural disorder in breast tissues. Malignant samples exhibited significantly higher IPR values and greater standard deviations compared to benign counterparts, indicating elevated mass density fluctuations and structural disorganization. This structural disorder was found to scale linearly with the disorder strength $L_d \sim \langle dn \rangle \cdot l_c$, further confirming the biophysical changes associated with malignancy.

These results demonstrate that fractal dimension metrics, functional distributions, and IPR-based disorder quantification provide powerful, complementary parameters for distinguishing malignant from benign breast tissue. This integrative, physics-based framework offers a promising pathway for developing non-invasive, image-based diagnostic tools for early breast cancer detection and grading.

**The X-ray mammogram data site:**

The dataset is available at **https://doi.org/10.17632/ywsbh3ndr8.2**

**Authors Contributions**:
PP conceived the idea and project; SM performed the main simulations, and MA helped him. PP wrote the first draft, SM and MA further developed it, and all authors contributed to the final version of the paper.

**Acknowledgments**
We acknowledge NIH and MSU for funding the project.

**Funding**
This work was partially supported by the National Institutes of Health under grant R21 CA260147 and ORED, Mississippi State.

**Data Availability Statement**

Data are available from the corresponding author upon reasonable request.

**Conflicts of Interest**

The authors declare that they have no conflicts of interest.